\documentclass[]{aa}
\usepackage{dcolumn}
\usepackage{hyperref}
\usepackage{graphicx}
\usepackage{subfigure} 
\usepackage{mathpazo,calc,tikz}
\usepackage{microtype}
\usepackage{array,booktabs,colortbl}
\usepackage{multirow}
\usepackage[version=3]{mhchem} 

\usepackage{savesym}
\usepackage{amsmath}
\savesymbol{iint}
\savesymbol{iiint}
\usepackage{txfonts}
\restoresymbol{TXF}{iint}
\restoresymbol{TXF}{iiint}

\begin{document}

\title{Tunnelling dominates the reactions of hydrogen atoms with unsaturated alcohols and aldehydes in the dense medium}
\author{V. Zaverkin\inst{\ref{inst1}} \and T. Lamberts\inst{\ref{inst1}}\inst{\ref{inst2}} \and M. N. Markmeyer\inst{\ref{inst1}} \and J. K\"{a}stner\inst{\ref{inst1}}}

\institute{Institute for Theoretical Chemistry, University Stuttgart, Pfaffenwaldring 55, 70569 Stuttgart, Germany \label{inst1} \and 
Current Address: Leiden Institute of Chemistry, Gorlaeus Laboratories, Leiden University, P.O. Box 9502, 2300 RA Leiden, The Netherlands, \email{a.l.m.lamberts@lic.leidenuniv.nl} \label{inst2}}

\date{Received XX/XX/XXXX / Accepted XX/XX/XXXX}

\abstract{ Hydrogen addition and abstraction reactions play an important role as surface reactions in the buildup of complex organic molecules in the dense interstellar medium. Addition reactions allow unsaturated bonds to be fully hydrogenated, while abstraction reactions recreate radicals that may undergo radical-radical recombination reactions. Previous experimental work has indicated that double and triple C--C bonds are easily hydrogenated, but aldehyde -C=O bonds are not. Here, we investigate a total of 29 reactions of the hydrogen atom with propynal, propargyl alcohol, propenal, allyl alcohol, and propanal by means of quantum chemical methods to quantify the reaction rate constants involved.

First of all, our results are in good agreement with and can explain the observed experimental findings. The hydrogen addition to the aldehyde group, either on the C or O side, is indeed slow for all molecules considered. Abstraction of the H atom of the aldehyde group, on the other hand, is among the faster reactions. Furthermore, hydrogen addition to C--C double bonds is generally faster than to triple bonds. In both cases, addition on the terminal carbon atom that is not connected to other functional groups is easiest. Finally, we wish to stress that it is not possible to predict rate constants based solely on the type of reaction: the specific functional groups attached to a backbone play a crucial role and can lead to a spread of several orders of magnitude in the rate constant.
}


\titlerunning{Hydrogen atom tunnelling dominates reactions with unsaturated compounds}
\authorrunning{Zaverkin et al.}

\maketitle

\section{Introduction}

In the dense molecular clouds of the interstellar medium, the low temperatures (10 - 20~K) result in the freeze-out of atoms and molecules on the surface of dust grains. There they can diffuse and finally react with each other. In particular, hydrogenation reactions of the type \ce{H + A -> HA} that result in the formation of fully saturated species can be facilitated by the uptake of the excess energy by a third body, the dust grain. For instance, the hydrogenation of the triply bonded carbon monoxide molecule, CO, leads to the formation of first formaldehyde, \ce{H2CO}, and finally methanol, \ce{CH3OH} \citep{Tielens:1982,Hiraoka:1998,Watanabe:2002,Fuchs:2009}. Carbon-carbon double and triple bonds can also be hydrogenated, starting from acetylene, \ce{C2H2}, leading to ethylene, \ce{C2H4}, and ethane, \ce{C2H6} \citep{Bennet:1973,Kobayashi:2017}. 

By adding more carbon and oxygen atoms to the backbone, this leads to the study of complex organic molecules (COMs). For instance, not only methanol, but also acetaldehyde, vinyl alcohol (tentative), and ethanol have been detected \citep{Ball:1970,Gottlieb:1973,Turner:2001,Zuckerman:1975}. However, in the variety of \ce{C3H_nO} species, the unsaturated molecules, propynal, propenal, and propanal, have indeed been detected \citep{Irvine:1988,Hollis:2004}, but propanol has not (yet) been observed. Various other molecules with three carbon atoms and one oxygen atom have also been seen, such as propylene oxide, acetone, and cyclopropenone \citep{McGuire:2016,Lykke:2017,Hollis:2006}. 

It is now well-known that methanol is mainly formed on the surfaces of dust grains through successive hydrogenation of carbon monoxide. Implementing these routes in models indeed allows reproducing the observed abundances \citep{Boogert:2015, Walsh:2016,Deokkeun:2017}. However, for acetaldehyde and ethanol, surface hydrogenation reactions alone do not seem to explain the relative abundances in young stellar objects \citep{Bisschop:2008}. Furthermore, \citet{Loison:2016} suggested that both propynal and cyclopropenone are exclusively formed in the gas phase. In an attempt to be able to link the various unsaturated \ce{C3H_nO} molecules to propanol, which is fully saturated, \citet{Jonusas:2017} performed hydrogenation experiments at low temperature under ultra-high vacuum conditions. They found that under their conditions and for a total H fluence of 2.7$\times 10^{18}$ cm$^{-2}$ s$^{-1}$, the aldehyde group, -HC=O, could not be reduced to the alcohol form. This may seem to be in line with the non-detection of propanol. However, it should be kept in mind that this only considers one reaction route.  Moreover, while laboratory experiments are very useful in determining reaction pathways, it is difficult, if not impossible, to determine exact reaction rate constants that can be used in astrochemical models. These rate constants are highly needed in order to determine which pathways dominate the formation of certain molecules.

Here, we further characterise the hydrogenation reactions starting from propynal and successively leading to propanol. We take a computational approach and make use of density functional theory benchmarked against coupled cluster theory in combination with instanton theory to quantify the various reaction rate constants involved. Moreover, not only hydrogen addition reactions are possible, but the degree of saturation can also decrease as a result of hydrogen abstraction reactions. Both types of reactions are studied here, leading to a total of 29 reactions under consideration. 

Section~\ref{sec:meth} gives an overview of the computational methods used, both the underlying electronic structure calculations and the basics of instanton theory. In Section~\ref{sec:res} the molecules under study are introduced along with the main reaction paths, subsequently, the calculated rate constants are discussed in comparison to the experimental work of \citet{Jonusas:2017}. Finally, we provide astrochemically relevant conclusions in Section~\ref{sec:con}.


\section{Methods}\label{sec:meth}
\subsection{Electronic structure}
The potential energy surface was described by density functional theory (DFT).
Following the benchmark calculations performed by \citet{Kobayashi:2017}, the
MPWB1K functional \citep{Zhao:2004} in combination with the basis set
def2-TZVP \citep{Weigend:1998} was chosen. The energy and gradient
calculations were carried out in NWChem version 6.6 \citep{Valiev:2010}.
An additional benchmark was performed for the activation and reaction energies of the four hydrogen addition reactions 
of propenal by comparing to single-point energy calculations at CCSD(T)-F12/VTZ-F12 level \citep{Knowles:1993,Knowles:2000,Deegan:1994,Knizia:2009,Adler:2007, Peterson:2008} in Molpro version 2012 \citep{MOLPRO:2012}.
For an accurate calculation of the rate constant, a good description of the barrier region is crucial. 
The differences between coupled cluster and DFT activation energies are only $\pm2$~kJ~mol$^{-1}$; see Appendix \ref{app:app_1}. 
We note that \citet{Alvarez-Barcia:2018} showed that the multi-reference
character for H addition to a C=O functional group is small and CCSD(T)-F12
values can be used as a reference.

The general procedure is the same for each selected molecule. Geometry optimisations are carried out for the separated
reactant, product, and transition structures and verified by the appropriate number of imaginary frequencies. 
A transition structure is characterised by the Hessian 
bearing exactly one negative eigenvalue. To confirm that the found transition
structure connects the desired reactant and {product} an 
intrinsic reaction coordinate (IRC) search is conducted. From the end-point of the IRC, a re-optimisation is performed
to obtain the pre-reactive complex (PRC).

All calculations were performed with DL-find \citep{Kaestner:2009}
within Chemshell \citep{Sherwood:2003, Metz:2014}. IRC searches were performed using the algorithm
described by \citet{Meisner:2017} and \citet{Hratchian:2004}.

\subsection{Reaction rate constants}
Reaction rate constants were calculated using instanton theory \citep{Langer:1967, Miller:1975, Callan:1977, Coleman:1977}, 
which has been proven to provide accurate tunnelling rates down to very low temperatures \citep{Kaestner_b:2014, Meisner_b:2016}. Instanton theory treats the quantum effects 
of atomic movements by Feynman path integrals. The instanton path can be seen as a periodic orbit in the upside-down potential connecting the reactant and product states at a given temperature. This path is located by
a Newton--Raphson optimisation scheme. More details on our implementation of instanton theory are given in \citet{Rommela:2011}, \citet{Rommelb:2011} and \citet{McConnell:2017}. Reactant partition functions in the bimolecular cases were calculated as products of the partition functions of both separated reactants. For the unimolecular case, the partition function of the respective
PRC was used. To model the reaction on a surface, the rotational partition function in the unimolecular case is assumed to be constant during the reaction. 
Such an implicit surface model \citep{Meisner_b:2017} covers the effect of the ice surface, where the rotation on the surface is suppressed. {In principle, this approximation holds as long as the interaction between the surface and the admolecules is weak, as is the case for a CO-rich ice mantle and the \ce{C3H_nO} molecules studied here.} The instanton calculations were performed in a stepwise cooling scheme, starting with instanton discretisation by 40 images and increasing the number of images for lower temperatures, if necessary. The initial temperature was chosen just below the crossover temperature $T_c$ \citep{Gillan:1987}, since canonical instanton theory is not applicable above it. The crossover temperature is defined as
\begin{equation}
T_c = \frac{\hbar \omega_b}{2 \pi k_B}
,\end{equation} 
where $\omega_b$ is the absolute value of the imaginary frequency at the transition state, $\hbar$ is the reduced Planck constant,
and $k_{\text{B}}$ is the Boltzmann constant. It provides an estimate of the temperature below which tunnelling dominates the rate constant. Instantons were optimised 
to a residual gradient below $10^{-8}$ au. 
A path of the transition structure elongated along the unstable mode was chosen as the initial guess for the calculation of the first instanton of each reaction. 
For every subsequent cooling step, the geometry of the preceding instanton was used as the initial configuration for the optimisation. Optimised instantons can be used to calculate both bi- and unimolecular rate constants by using the respective reactant partition function. In this work we focus on unimolecular rate constants, since these are related to the Langmuir--Hinshelwood surface reaction mechanism that is relevant at the low temperatures present in dense clouds in the ISM. 
All instantons were calculated down to the temperature of 60 K. 

\section{Results}\label{sec:res}

We first introduce the five investigated molecules and their possible reactions along with DFT calculations of the activation and reaction energies. Subsequently, we discuss the calculated rate constants for all exothermic reactions. Finally, we relate the calculated rate constants to previous experimental observations \citep{Jonusas:2017}.

\subsection{Selected molecules}
\label{sec:sec_31}
\begin{figure}[t!]
\begin{center}
\includegraphics[scale=0.23]{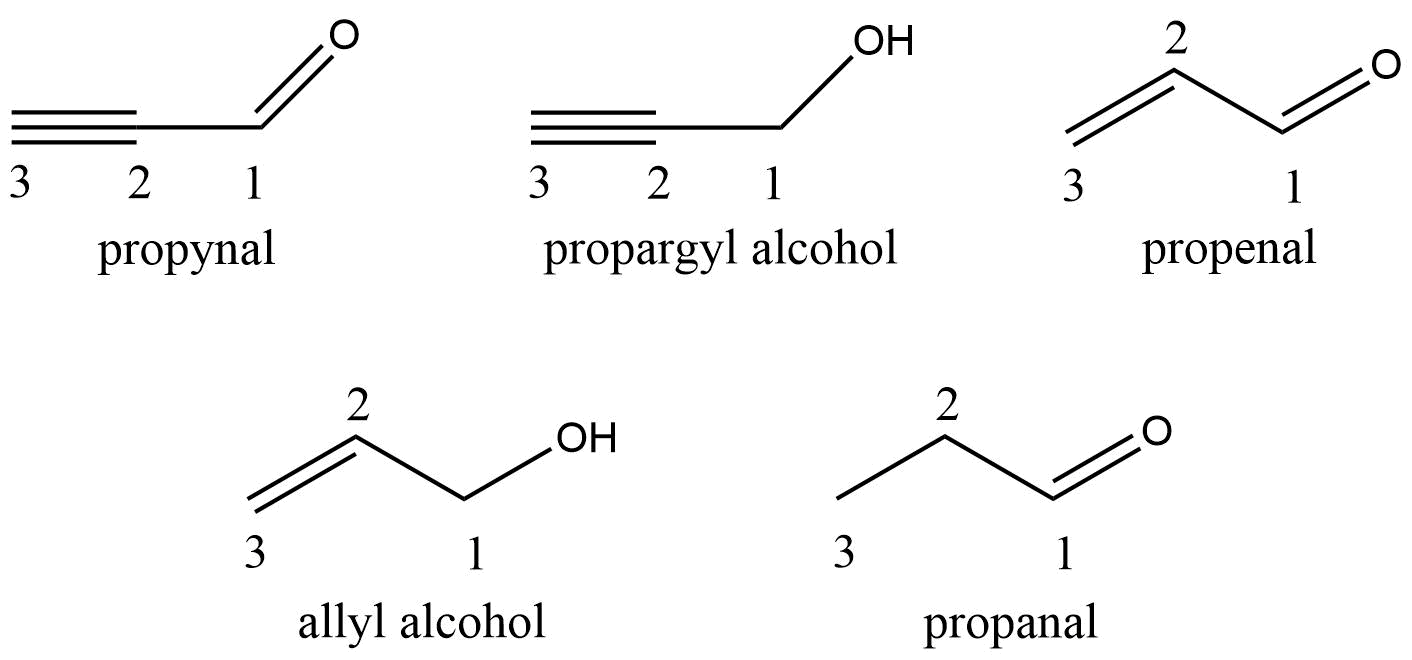}
\caption{Investigated molecules, containing carbon double
  or triple bonds, and aldehyde or alcohol groups: propynal, propargyl
  alcohol, propenal, allyl alcohol, and propanal.  Carbon atoms are enumerated
  starting from the one nearest to the functional group.}
\label{fig:fig1}
\end{center}
\end{figure}

We performed calculations for hydrogen addition and abstraction reactions of five unsaturated organic molecules, each containing an {aldehyde group, alcohol group, carbon double or triple bond}. These are, starting from the most 
unsaturated molecule, propynal, propargyl alcohol, propenal, allyl alcohol, and 
propanal. In order to simplify the nomenclature of 
possible reaction paths, each carbon atom of the selected molecules is labelled as depicted in Fig. \ref{fig:fig1}.

The mass of the migrating particles, the height, and the width of the barrier are three important parameters that have impact
on quantum tunnelling. Since all reactions we study here are hydrogen addition or abstraction reactions, the effective mass can be considered to be similar in all cases.
Activation and reaction energies excluding ($E_{\text{A}}$ and $E_{\text{r}}$, respectively) and including ($E_{\text{A}}^{0}$ and $E_{\text{r}}^{0}$,
respectively) zero-point vibrational energy (ZPE) were calculated relative to
the PRC. They are given, along with the crossover temperatures, in Table
\ref{tab:tab1}.

\begin{table}[t!]
\begin{center}
  \caption{Activation energies $E_{\text{A}}$ and reaction energies
    $E_{\text{r}}$ of hydrogen addition and abstraction reactions for
    the investigated molecules in kJ~mol$^{-1}$ relative to the
    PRC.  Values including ZPE are denoted $E_{\text{A}}^{0}$ and
    $E_{\text{r}}^{0}$. $T_{\text{c}}$ is the
    crossover temperature in K.}
\label{tab:tab1}
\begin{tabular}{lrrrrr}
\toprule
Reaction & $E_{\text{A}}$ & $E_{\text{A}}^{0}$ &
 $E_{\text{r}}$ & $E_{\text{r}}^{0}$& $T_{\text{c}}$\\
\midrule \midrule
\multicolumn{4}{l}{Propynal}\\
\midrule
add C$_1$&      21.9&   24.2&   $-$111.5&        $-$88.0&       219\\  
add C$_2$&      18.5&   18.9&   $-$183.6&       $-$159.5&       175\\ 
add C$_3$&      11.2&   12.6&   $-$222.4&       $-$197.9&       160\\ 
add O    &      29.6&   31.0&   $-$202.8&       $-$177.5&       297\\ 
abs C$_1$&      27.3&   20.0&    $-$60.2&        $-$69.3&       353\\   
abs C$_3$ $^{a}$ &      --  &    -- &      139.7&          134.2&        --\\   
\midrule \midrule
\multicolumn{4}{l}{Propargyl alcohol}\\
\midrule
add C$_2$            & 16.9&17.8&$-$179.0&$-$154.1&169\\
add C$_3$& 14.3&15.0&$-$190.4&$-$165.9&177\\
abs C$_1$& 25.3&17.7&$-$80.6&$-$92.1&326\\
abs C$_3$ $^{a}$& -- & -- & 141.7&131.6& --\\
abs O     & 59.5&50.9&2.0&$-$10.1&376\\
\midrule \midrule
\multicolumn{4}{l}{Propenal}\\
\midrule 
add C$_1$& 25.7&26.9& $-$90.5&$-$67.6&228\\
add C$_2$& 16.5&18.0& $-$152.6&$-$132.2&181\\
add C$_3$& 7.0&8.2& $-$212.3&$-$188.6&127\\
add O    & 32.3&33.0& $-$186.7&$-$162.9&317\\
abs C$_1$& 24.9&17.3& $-$47.6&$-$55.4&329\\
abs C$_2$& 74.2&65.0& 48.8&36.4&235\\
abs C$_3$& 75.3&67.0& 46.9&35.1&248\\
\midrule \midrule
\multicolumn{4}{l}{Allyl alcohol}\\
\midrule 
add C$_2$& 14.9&16.5& $-$169.4&$-$148.6&171\\
add C$_3$& 7.5&9.0& $-$178.8&$-$157.8&138\\
abs C$_1$& 33.0&24.5& $-$92.9&$-$104.6&328\\
abs C$_2$& 63.2&54.6& 33.0&21.1&265\\
abs C$_3$& 71.6&63.2& 44.0&32.1&247\\
abs O    & 56.9&46.1& 5.6&$-$7.5&400\\
\midrule  \midrule
\multicolumn{4}{l}{Propanal}\\
\midrule
add C$_1$& 21.8&23.9& $-$104.5&$-$82.6&206\\
add O& 39.8&41.1& $-$136.6&$-$110.1&335\\
abs C$_1$& 21.3&13.8& $-$55.1&$-$63.0&309\\
abs C$_2$& 33.9&25.4& $-$54.3&$-$64.6&350\\
abs C$_3$& 53.9&44.8& 6.5&$-$8.7&316\\
\bottomrule
\multicolumn{6}{l}{$^{a}$ {Energy relative to the separated reactants.}}\\
\end{tabular}
\end{center}
\end{table}

The barriers for hydrogen addition reactions, including zero-point energies, lie between 8.2~kJ~mol$^{-1}$ and 
41.1~kJ~mol$^{-1}$. The highest activation energies were observed for the hydrogen addition on the O atom of the aldehyde group in accordance with results obtained for glyoxal and glycoaldehyde by \citet{Alvarez-Barcia:2018}. The lowest barriers are found for hydrogen additions to the C$_3$ atom of either a double or triple bond.
Activation energies for the hydrogen abstraction reactions are generally somewhat higher than those for hydrogen addition reactions and range from 13.8~kJ~mol$^{-1}$ to 67.0~kJ~mol$^{-1}$. The hydrogen abstraction reaction with the highest activation energy is the one at the C$_3$ 
position for all molecules, while abstraction at C$_1$ is comparatively easier. The C$_3$ hydrogen abstraction reactions of propynal and propargyl alcohol are highly endothermic (on the order of 130~kJ~mol$^{-1}$), and thus are not considered in the following. A PRC structure could not be found for hydrogen abstraction at the C$_3$-atom from propargyl alcohol. Thus, we used
the separated molecules as reference. 

For moderately endothermic reactions, like hydrogen abstraction from C$_2$ and C$_3$ for propenal and allyl alcohol, no rate constants were calculated since they are very low at low-temperature conditions. Tunnelling can be expected to only occur through the exothermic fraction of the barrier.  The three reactions that become exothermic when including zero-point energy, on the other hand, are also discussed in
the following sections.

\subsection{Rate constants}
All calculated rate constants here concern reactions between closed-shell molecules and hydrogen atoms. The resulting radicals can subsequently react with another radical, for example\emph{}, a hydrogen atom through generally barrier-less, radical-recombination reactions. Thus, in the overall process, the rate constants we provide here are rate limiting, unless the orientation of the reactants prevents reactions from taking place.

\begin{figure*}[h!]
\centering
\subfigure[]
{\includegraphics[width=0.46\textwidth]{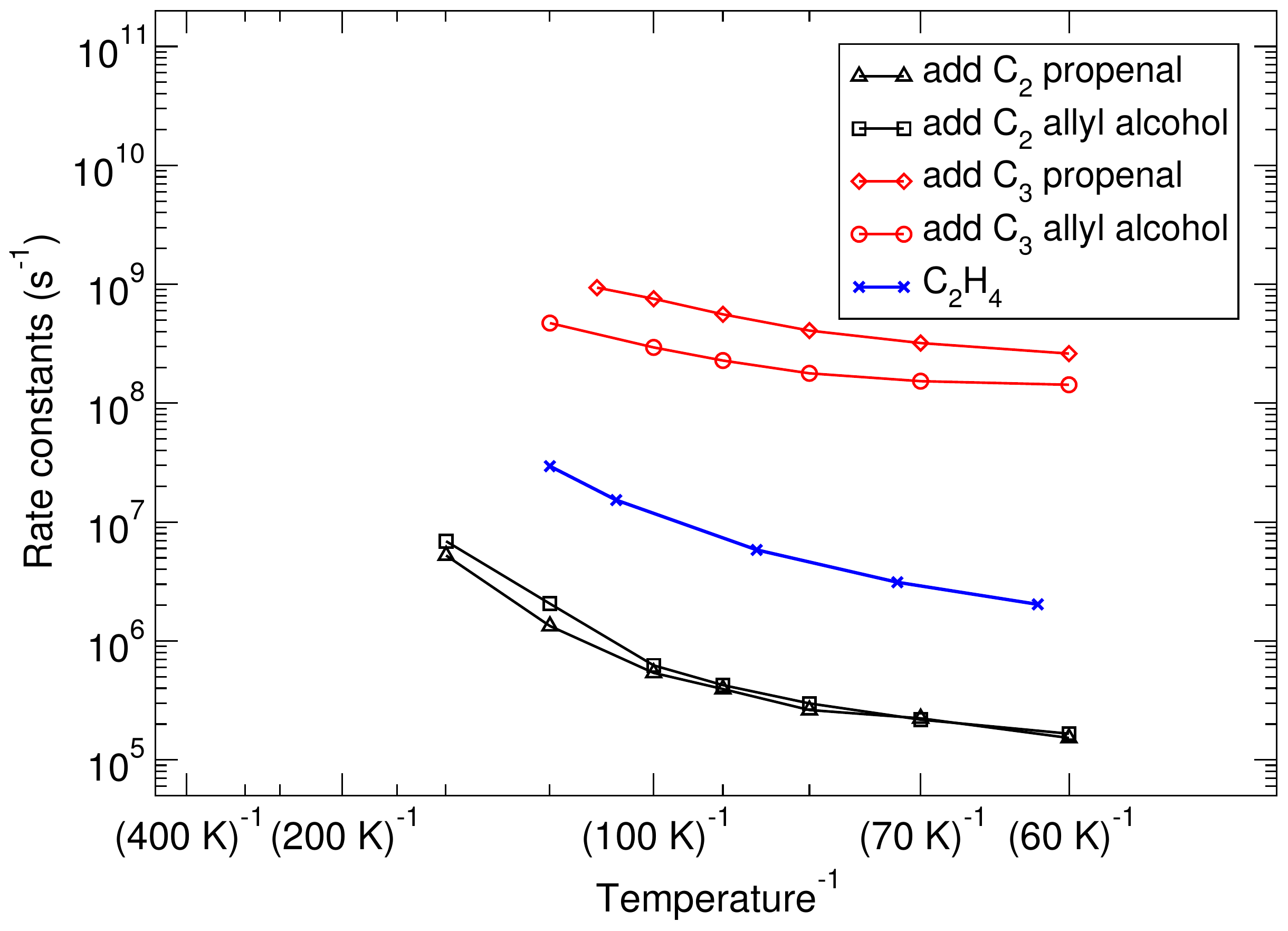}}
\subfigure[]
{\includegraphics[width=0.46\textwidth]{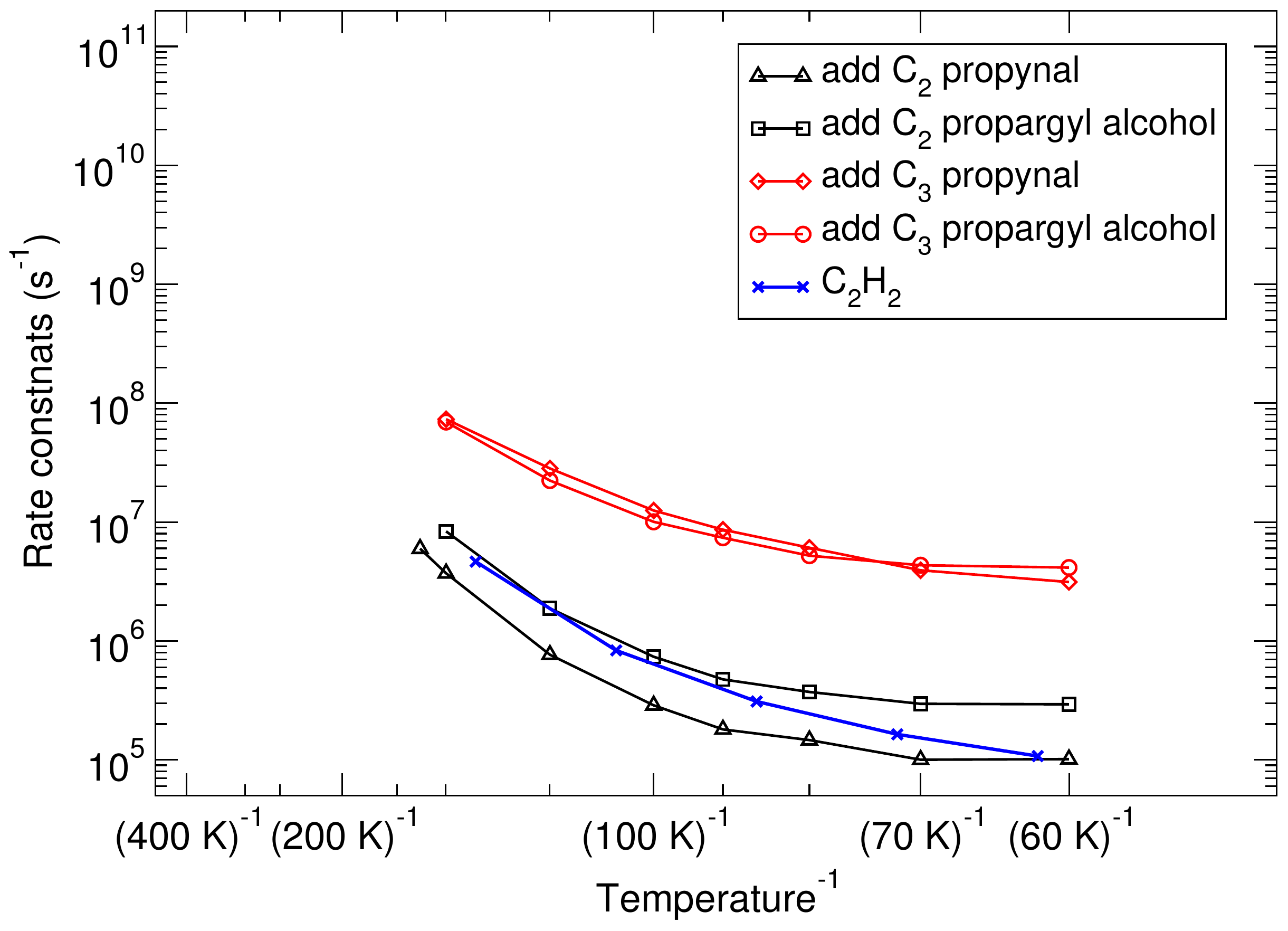}}
\subfigure[]
{\includegraphics[width=0.46\textwidth]{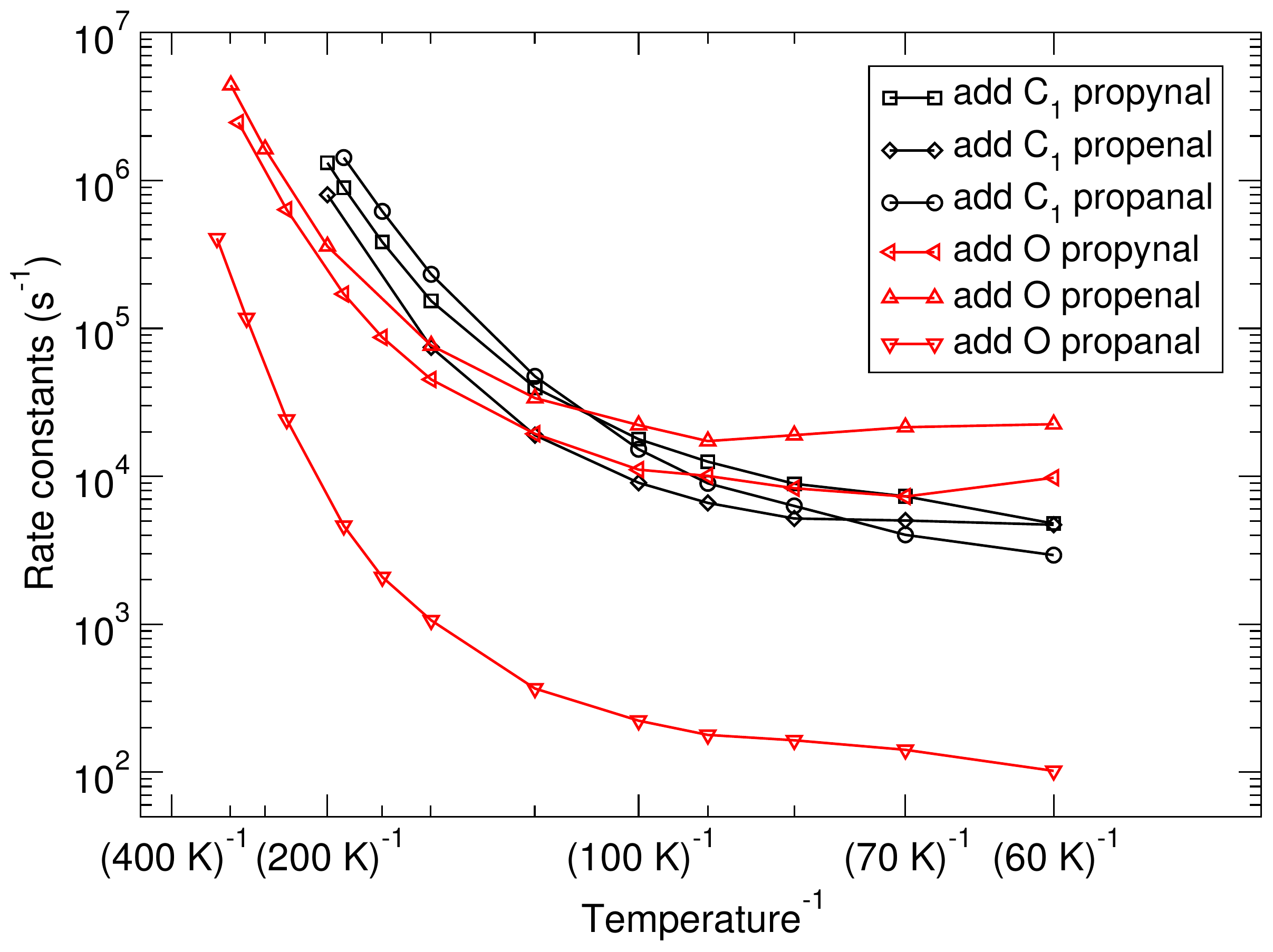}}
\subfigure[]
{\includegraphics[width=0.46\textwidth]{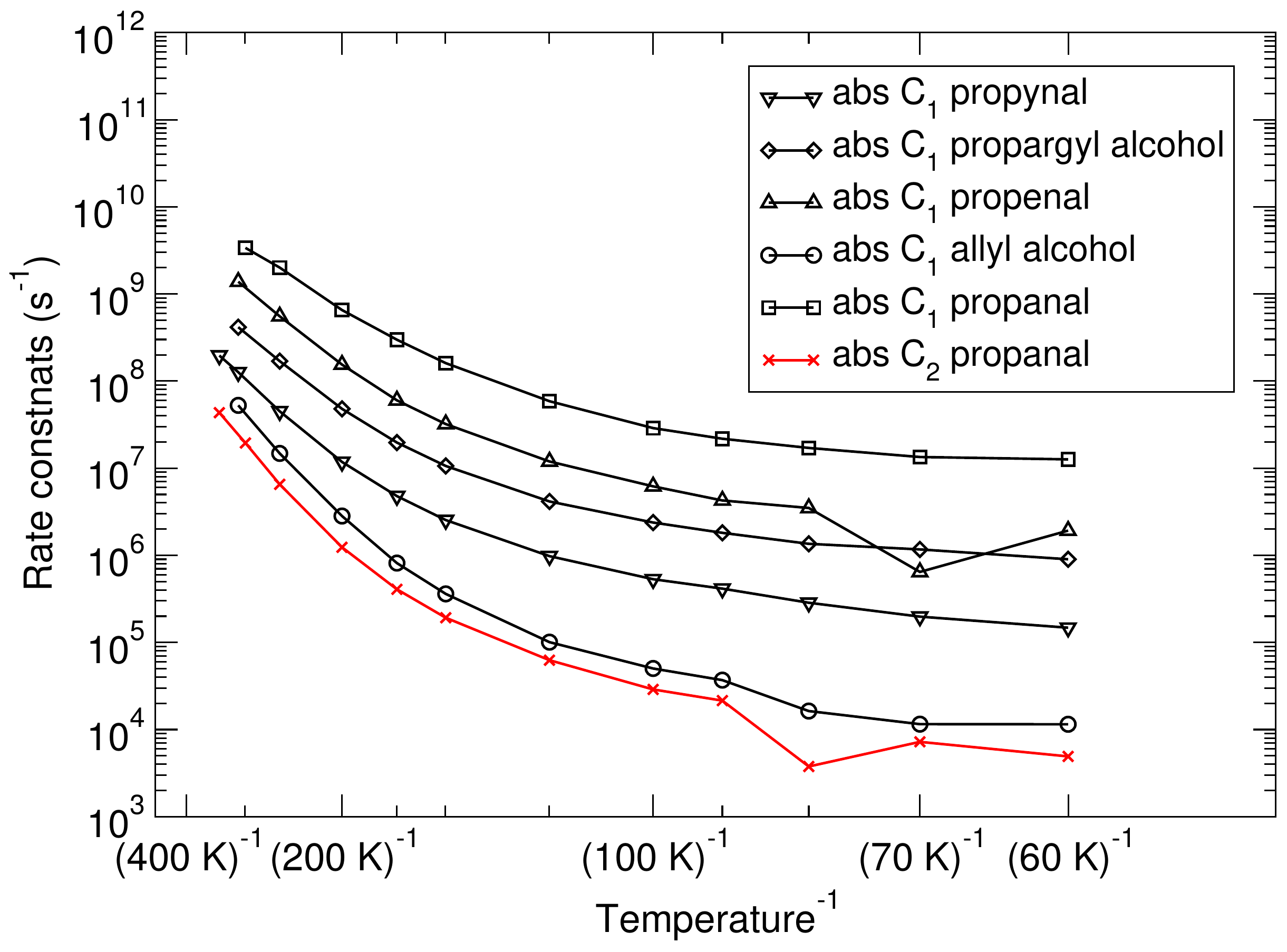}}
\subfigure[]
{\includegraphics[width=0.46\textwidth]{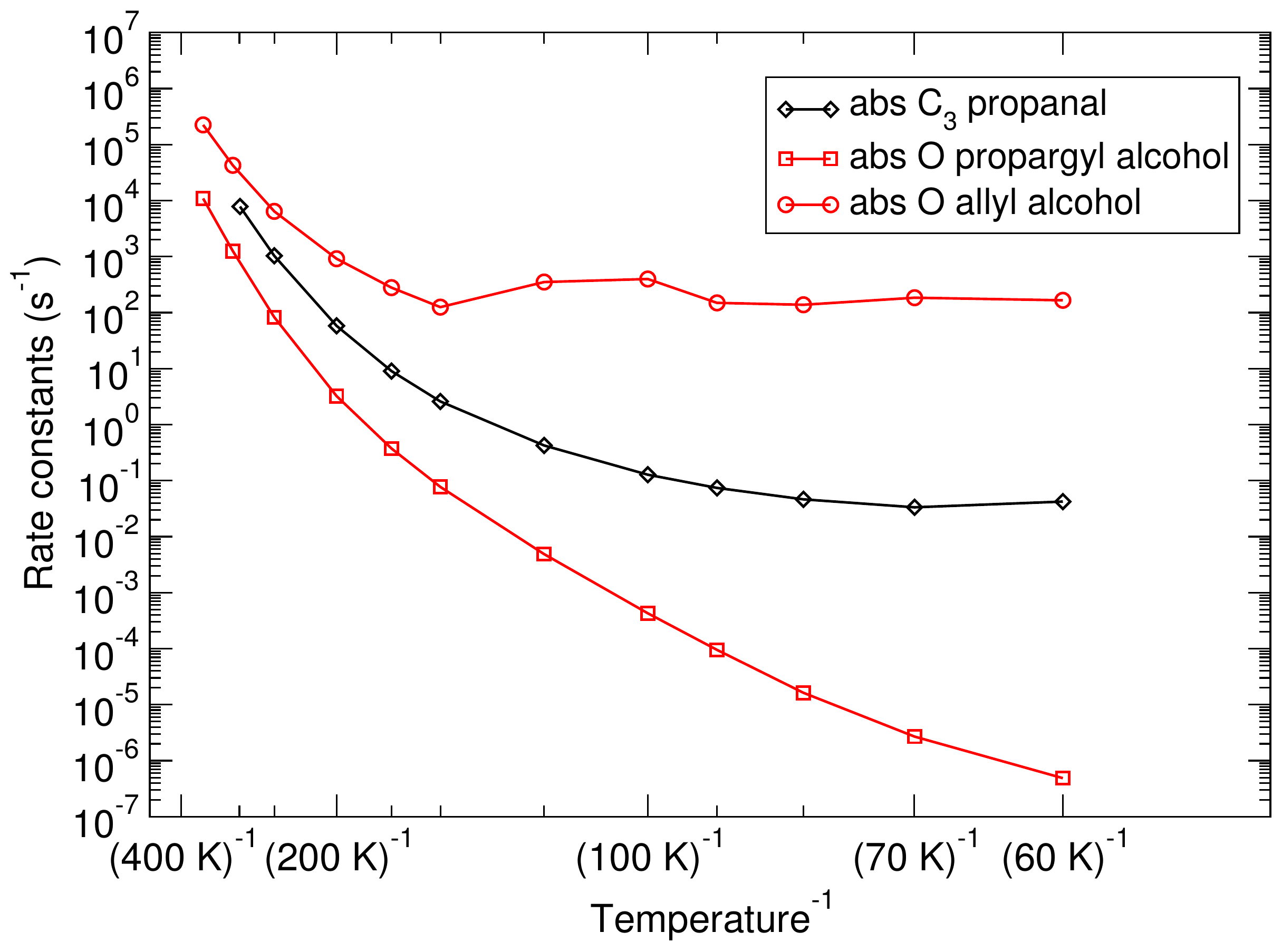}}
\caption{Instanton rate constants for the reactions of hydrogen atoms with the investigated molecules. (a) Hydrogen additions to carbon double bonds. Values for C$_2$H$_2$
and C$_2$H$_4$ are taken from \citet{Kobayashi:2017}. (b) Hydrogen addition to carbon triple bonds. (c) Hydrogen additions to aldehyde groups. (d) Hydrogen abstractions from C$_1$ and C$_2$. (e) Hydrogen abstractions from C$_3$ and O (exothermic due to ZPE).}
\label{fig:fig_2}
\end{figure*}

The temperature-dependent unimolecular rate constants calculated below the crossover-temperature by canonical instanton theory are given in Fig. \ref{fig:fig_2} for both the hydrogen addition and hydrogen abstraction reactions. {We note that the deviation from the smooth curves visible in panel (d) for instance is thought to be due to numerical errors, and these irregularities are thus not physically meaningful.} For numerical values, we refer to Appendix \ref{app:app_2}. The plots are ordered according to the bond type, that is, addition to carbon double and triple bonds in Fig.~\ref{fig:fig_2}(a) and (b), addition to the aldehyde C and O atoms in panel (c), and abstraction reactions in panels (d) and (e). 
The rate constants for different reaction paths range over almost 16 orders of magnitude; the y-axis scales in Fig. \ref{fig:fig_2} vary. 

The rate constant does not only depend on the barrier height, but also on the shape of the barrier. The easiest estimation for that is the curvature of the classical transition state \citep{Gillan:1987}, which can be extracted from the crossover temperatures given in Table~\ref{tab:tab1}. A high crossover temperature indicates a thin barrier. The full IRC paths for the reactions of the hydrogen atoms with the investigated molecules are presented in Appendix \ref{app:app_irc}. We note that at low impact energies, that is\emph{}, when the temperature is low, {the shape of the barrier along the IRC may be distinctly different from that at the classical transition state} (see Fig. \ref{fig:fig2}).

Comparing the hydrogen addition reactions to the C$_3$ atoms for the double and triple bonds, Fig.~\ref{fig:fig_2}(a) and~\ref{fig:fig_2}(b), we find that addition to the double bonds is faster than to the triple bonds, in accordance with results obtained for C$_2$H$_2$ and C$_2$H$_4$ \citep{Kobayashi:2017}. Moreover, addition to the C$_3$ atom is generally the fastest reaction pathway compared to the other possibilities within the same molecules. This in turn is in agreement with the activation energies presented in Table~\ref{tab:tab1}, since these reactions have the lowest energy barriers. 

The rate constants for the hydrogenation of the aldehyde group show the lowest reactivity, as depicted in Fig.~\ref{fig:fig_2}(c), regardless of whether the addition takes place at the carbon or oxygen atom.  
Although the barriers for addition to O atoms are higher, they are somewhat narrower (except for propanal), as can be seen in the figure in the appendix, Fig. \ref{fig:fig2}(c). This leads to the observed crossing of the rate constants around 100~K. In the high-temperature regime, the addition to O is slower, but at lower temperatures, the barrier width dominates the tunnelling behaviour, as explained above, and indeed the hydrogen addition to O surpasses the addition to C$_1$. 
The hydrogenation of the oxygen atom of propanal has a much higher and therefore broader barrier than the other molecules, leading to much lower rate constants.

The hydrogen abstraction rate constants vary strongly among the selected molecules, as can be seen in Fig.~\ref{fig:fig_2}(d) and Fig.~\ref{fig:fig_2}(e). The rate constants for hydrogen abstraction from C$_1$  reversely follow the activation energy and crossover temperature values presented in Table \ref{tab:tab1}. In other words, the instanton rate constants run parallel to each other, but their values are shifted according to the barrier height, since the barrier shape is very similar. 
The three abstraction reactions that are exothermic only when considering the ZPE have very similar barrier heights, resulting in similar rate constants at high temperature. At low temperature, especially the abstraction of a hydrogen atom from the alcohol group of propargyl alcohol is notable, where the rate constant decreases steeply with decreasing temperature as a result of its much wider barrier at low impact energies (see Fig.~\ref{fig:fig2}(e)).

From the above it becomes clear that it is not possible to extrapolate rate constants from small test systems directly to other molecules that may contain the same bond type, but different functional groups. The easiest example to demonstrate this is the H-addition to carbon double or triple bonds, where previous calculations on \ce{C2H2} and \ce{C2H4} differ by up to two orders of magnitude from those calculated here for the \ce{C2H_nO} species. Furthermore, the hydrogen abstraction reactions also clearly show that although the barrier shapes are very similar, the barrier height differs so much between the five molecules considered here that the rate constants also spread over three orders of magnitude.

\subsection{Comparison with experiments}
We briefly sketch the main experimental results per molecule as obtained by
\citet{Jonusas:2017} and comment how our calculated rate constants relate to
their experiments. We note that experiments were carried out at 10~K, whereas our
low-temperature rate constants are calculated at 60~K.  Overall, we find {good} agreement with the experimental data.

\textbf{Propynal.} \emph{\textup{Experiments show that the hydrogenation of propynal directly leads to the formation of propanal, leaving propenal as a short-lived species. Furthermore, hydrogen addition to the aldehyde group of the propynal was proposed not to lead to propargyl alcohol.}}\\
The calculated rate constants for the hydrogenation of the C$_3$ carbon atom of propenal are at least one order of magnitude higher than those for the hydrogenation of the C$_3$ carbon atom of propynal. This supports the experimental results of propenal as short-lived species. Addition at the C$_2$ atom or abstraction from the C$_1$ atom have lower values for the rate constant, and therefore we expect the addition to the C$_3$ carbon to be followed by a barrier-less addition to the C$_2$ carbon as a second step.
Finally, hydrogenation of the aldehyde group of propynal is indeed much slower than the hydrogenation of the carbon double or triple bonds and is therefore not competitive. 

\textbf{Propenal.} \emph{\textup{Experimentally, hydrogenation of propenal only leads to the reduction of the carbon double bond.}}\\ 
We find very similar results as for propynal, that is , the rate constant for addition to the aldehyde group is several orders of magnitude lower than for hydrogen addition to the C$_3$ carbon atom. Both addition to C$_2$ and abstraction from C$_1$ are slower as well, suggesting that again addition to the C$_3$ carbon is the dominant pathway.

\textbf{Propanal.} \emph{\textup{The experimental hydrogenation of propanal did not lead to propanol formation.}}\\ 
For the reaction of the hydrogen atom with propanal, the possibilities are addition to the aldehyde C or O atom or abstraction from C$_1$, C$_2$, or C$_3$. Out of these, the fastest reaction by more than three order of magnitude is abstraction from the C$_1$ carbon atom. This may lead to a cycle of H abstraction and subsequent barrier-less H addition reactions on the same carbon atom, leading to a net zero result. It is expected that because of such a cycle, indeed no propanol could be formed during the experiments.

\textbf{Allyl alcohol and propargyl alcohol.} \emph{\textup{Experimentally, the hydrogenation of propargyl alcohol ice leads mainly to allyl alcohol, although no further reaction to propanol is observed. Hydrogenation of a pure allyl alcohol sample, on the other hand, did lead to the formation of propanol. It was furthermore concluded that the activation energy is higher for the hydrogenation of the triple bond of propargyl alcohol than for the double bond of allyl alcohol. }} \\
Calculated rate constants and barrier relations are similar to those discussed above, that is\emph{}, faster hydrogenation for double bonds than for triple bonds. The C$_1$ hydrogen abstraction reaction from propargyl alcohol may compete with both the C$_2$ and C$_3$ hydrogen addition reactions, which can lead to an even slower reduction of the triple bond. Hydrogen abstraction from the OH group of the alcohols is slower than all competing reactions in both cases.

We wish to stress that experimental non-detection or the low values for the calculated rate constants do not imply that a reaction does not take place on interstellar timescales. This may in particular be important when specific reactive sites are blocked as a result of the orientation on an ice surface.

Finally, as for the formation of propanol in the solid state in the interstellar medium, hydrogenation of aldehydes with three carbon atoms does not seem to be an efficient route. However, other surface routes may still be available, such as the recombination of \ce{C2H_n} and \ce{CH2OH} fragments. 

\section{Conclusions}\label{sec:con}
Unimolecular reaction rate constants have been calculated and are provided for hydrogen addition and abstraction reactions from propynal, propargyl alcohol, propenal, allyl alcohol, and propanal and are thus available to be implemented in both rate-equation and kinetic Monte Carlo models aimed at studying the formation of COMs at low temperatures. Our results are generally in agreement with ultra-high vacuum experiments. 

Specifically, we can say that hydrogen addition to the C$_3$ atom is fastest, most notable for double bonds. However, addition to the C$_3$ atom for carbon triple bonded molecules is still faster than addition to either the C$_1$ or O atom of the aldehyde group.

For selected molecules, hydrogen abstraction reactions are in competition with addition reactions, overall slowing down the saturation of unsaturated bonds. However, on average, abstraction reactions have higher energy barriers and can also be endothermic in some cases.

Not only are hydrogen addition and abstraction in competition with each other, but the larger the molecules are, hydrogen diffusion starts to play a role as well in diffusion from on side of the molecule to another.

Finally, every molecule is unique, and therefore it is not wise to use small model systems to extrapolate rate constants for large molecules with similar bond types. Compare for instance the values for C$_2$H$_2$ and C$_2$H$_4$ versus those for the molecules studied here.

\section*{Acknowledgements}
This project was financially supported by the European Union’s Horizon 2020 research and innovation programme (grant agreement No. 646717, TUNNELCHEM) for VZ, MM, and JK, and the Alexander von Humboldt Foundation, the Netherlands Organisation for Scientific Research (NWO) via a VENI fellowship (722.017.008) and COST Action CM1401 via a STSM for TL.

\bibliographystyle{aa} 
\bibliography{viktor.bib}
\begin{appendix}
\section{Benchmark}
\label{app:app_1}

\begin{table}[h!]
\centering
\caption{Relative energies with respect to the separated reactants of the bimolecular hydrogen addition reactions for the propenal molecule without vibrational ZPE. Comparison of
CCSD(T)-F12 and DFT values. Energy values are given in kJ/mol.}
\begin{tabular}{lcc}
\toprule
Method & CCSD(T)-F12/ &MPWB1K/\\
      & VTZ-F12&def2-TZVP\\
\midrule \midrule
TS C$_1$& 25.7&25.5\\
TS C$_2$&14.4&16.3\\
TS C$_3$ & 6.0&6.8\\
TS O & 33.7&32.3\\
PS C$_1$&-74.6&-90.8\\
PS C$_2$&-148.8&-152.8\\
PS C$_3$& -199.1&-212.4\\
PS O& -176.9&-189.2\\
\bottomrule
\end{tabular}
\end{table}
\section{IRC: Classical reaction paths}
\label{app:app_irc}
\begin{figure*}[h!]
\centering
\subfigure[]
{\includegraphics[width=0.455\textwidth]{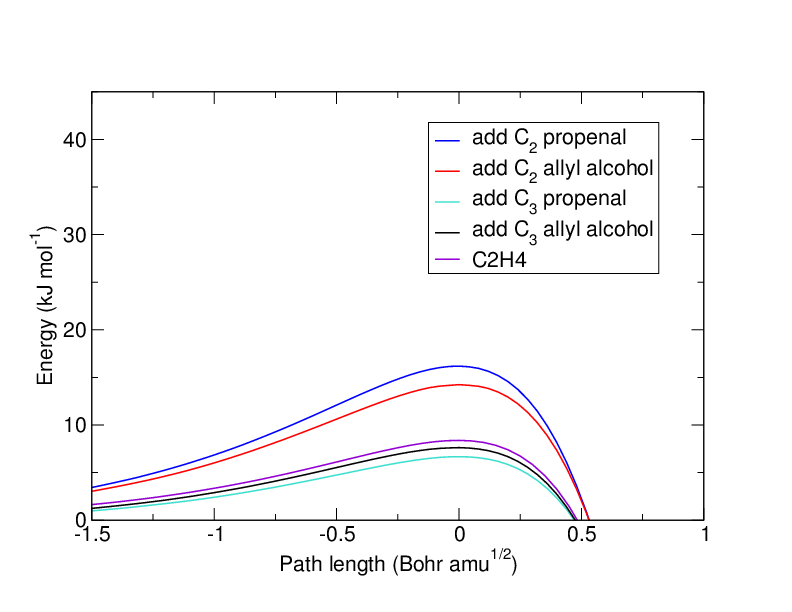}}
\subfigure[]
{\includegraphics[width=0.455\textwidth]{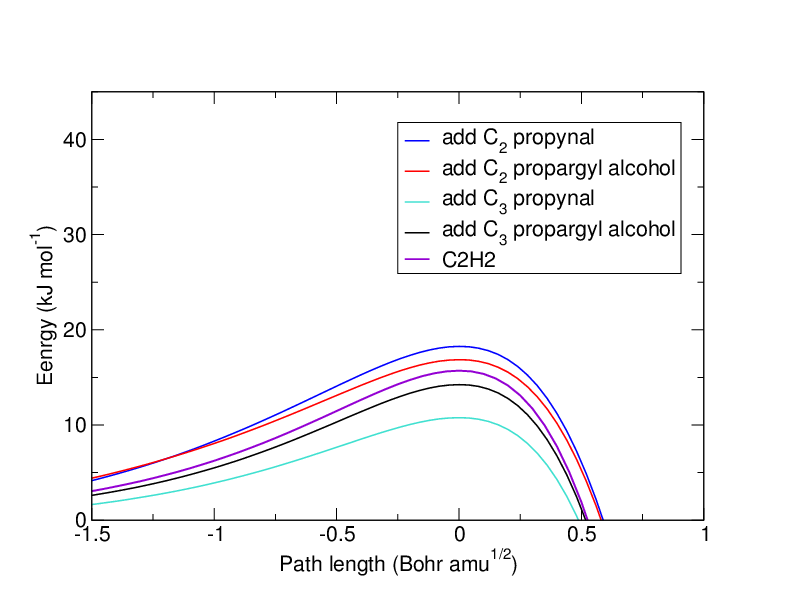}}
\subfigure[]
{\includegraphics[width=0.455\textwidth]{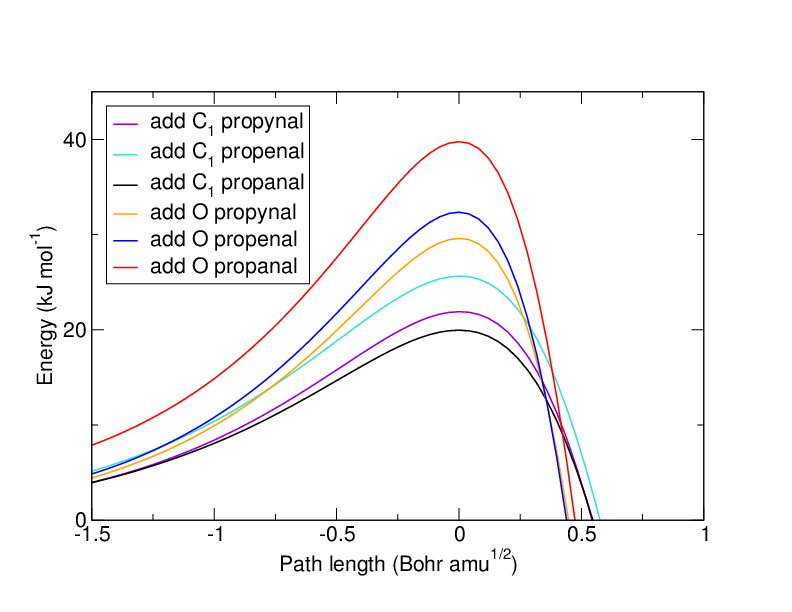}}
\subfigure[]
{\includegraphics[width=0.455\textwidth]{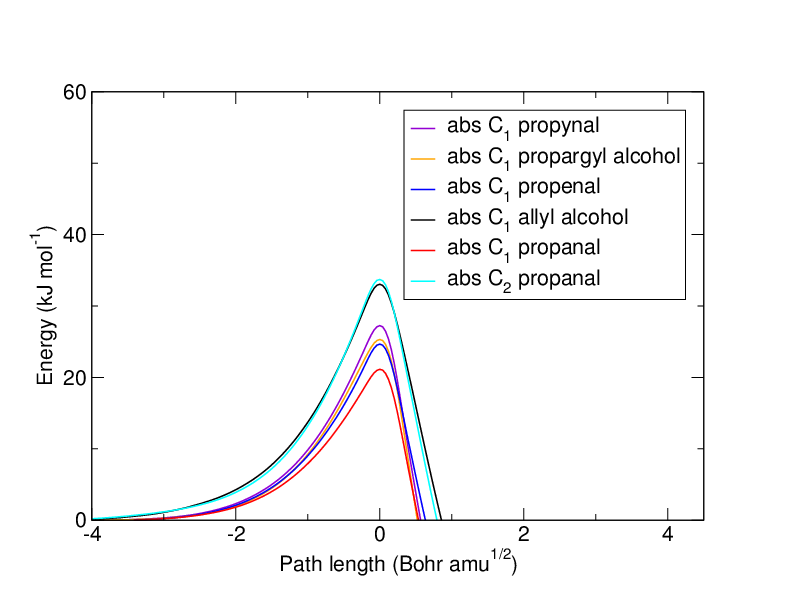}}
\subfigure[]
{\includegraphics[width=0.455\textwidth]{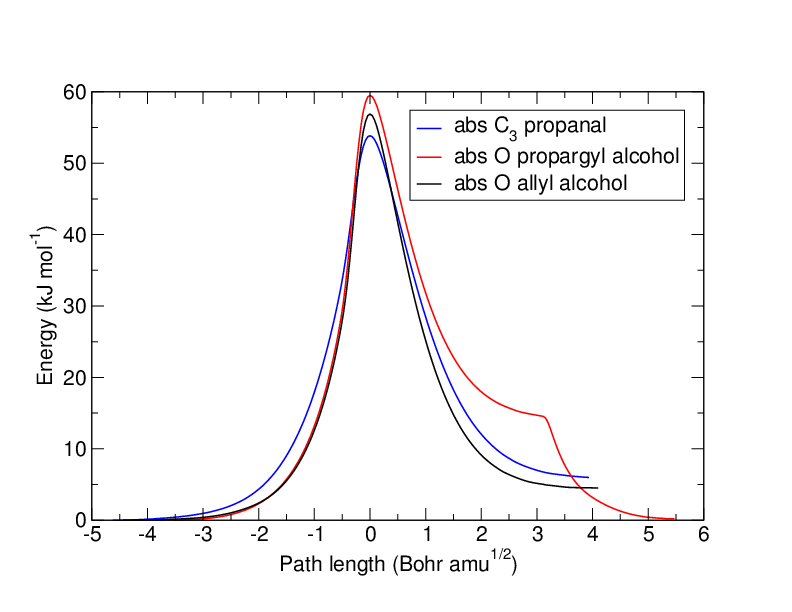}}
\caption{IRCs for the reactions of hydrogen atoms with the investigated molecules. (a) Hydrogen additions to carbon double bonds. (b)
Hydrogen additions to carbon triple bonds. (c) Hydrogen additions to aldehyde groups. Values for C$_2$H$_2$
and C$_2$H$_4$ are taken from \citet{Kobayashi:2017}. (d) Hydrogen abstractions from C$_1$ and C$_2$. (e)
Hydrogen abstractions from C$_3$ and O (Exothermic due to ZPE).}
\label{fig:fig2}
\end{figure*}

The IRC paths for the reactions of hydrogen atoms with the investigated molecules are presented in Fig. \ref{fig:fig2}. 
For the hydrogen addition reactions, IRCs are ordered according 
to the bond type. The barrier widths are similar for reactions proceeding
at the same bond type. Additionally, a comparison with the hydrogenation energy barriers of
C$_2$H$_2$ and C$_2$H$_4$ was carried out. Although the IRCs indeed follow a
similar trend qualitatively, quantitative, differences in the barrier height are present. 

For the hydrogen abstractions from the C$_1$ site, all reactions are exothermic. 
The hydrogen abstractions from the alcohol group for allyl and propargyl alcohol become exothermic 
when including zero-point energy, but have very high activation energies. For the hydrogen abstraction
from the alcohol group of propargyl alcohol, the additional 
structure rearrangement after the reaction contributes to the exothermicity. We note that the IRCs look quite
different in terms of the width, although the only difference in the molecule is the existence of a double or triple bond.
Again, this shows that it is not easily possible to extrapolate reaction energetics by reactions on the similar functional groups.
\section{Rate constant values}
\label{app:app_2}
\begin{table}
\parbox[t]{.45\linewidth}{
\centering
\caption{Instanton rate constants for the hydrogen addition to propynal. $k^{\text{bi}}$ is the bimolecular
rate constant given in cm$^3$s$^{-1}$. $k^{\text{uni}}$ is the unimolecular rate constant given in s$^{-1}$. $T$ is
the temperature given in K.}
\label{tab:tab3}
\begin{tabular}{ccc}
\toprule
T & $\log k^{\text{bi}}$ & $\log k^{\text{uni}}$ \\
\midrule \midrule
\multicolumn{3}{c}{C$_1$}\\
\midrule
200&    -16.64& 6.12\\
190&    -16.84& 5.95\\
170&    -17.28& 5.58\\
150&    -17.76& 5.19\\
120&    -18.51& 4.60\\
100&    -19.00& 4.25\\
90&     -19.24& 4.10\\
80&     -19.50& 3.95\\
70&     -19.71& 3.86\\
60&     -20.06& 3.68\\
\midrule\midrule 
\multicolumn{3}{c}{C$_2$}\\
\midrule
160&    -16.58& 6.78\\
150&    -16.83& 6.57\\
120&    -17.70& 5.88\\
100&    -18.30& 5.46\\
90&     -18.61& 5.26\\
80&     -18.83& 5.17\\
70&     -19.16& 5.00\\
60&     -19.37& 5.01\\
\midrule 
\midrule
\multicolumn{3}{c}{C$_3$}\\
\midrule 
150&    -14.49& 7.86\\
120&    -15.04& 7.45\\
100&    -15.51& 7.10\\
90&     -15.75& 6.94\\
80&     -15.98& 6.78\\
70&     -16.26& 6.60\\
60&     -16.48& 6.50\\
\midrule 
\midrule
\multicolumn{3}{c}{O}\\
\midrule 
280&    -15.95& 6.39\\
230&    -16.67& 5.80\\
190&    -17.37& 5.23\\
170&    -17.74& 4.94\\
150&    -18.11& 4.65\\
120&    -18.65& 4.28\\
100&    -19.05& 4.04\\
90&     -19.19& 4.00\\
80&     -19.39& 3.92\\
70&     -19.60& 3.86\\
60&     -19.66& 3.99\\
\bottomrule 
\end{tabular}
}
\hfill
\parbox[t]{.45\linewidth}{
\centering
\caption{Instanton rate constants for the hydrogen abstraction from propynal. $k^{\text{bi}}$
is the bimolecular
rate constant given in cm$^3$s$^{-1}$. $k^{\text{uni}}$ is the unimolecular rate constant given in s$^{-1}$. $T$ is
the temperature given in K.}
\label{tab:tab4}
\begin{tabular}{ccc}
\toprule
T & $\log k^{\text{bi}}$ & $\log k^{\text{uni}}$ \\
\midrule \midrule
\multicolumn{3}{c}{C$_1$}\\
\midrule
330&    -13.30& 8.29\\
300&    -13.55& 8.10\\
250&    -14.10& 7.65\\
200&    -14.81& 7.07\\
170&    -15.30& 6.68\\
150&    -15.65& 6.40\\
120&    -16.19& 5.99\\
100&    -16.57& 5.73\\
90&     -16.74& 5.62\\
80&     -16.97& 5.46\\
70&     -17.22& 5.30\\
60&     -17.44& 5.17\\
\midrule 
\\
\\
\end{tabular}
\centering
\caption{Instanton rate constants for the hydrogen addition to propargyl alcohol. $k^{\text{bi}}$
is the bimolecular
rate constant given in cm$^3$s$^{-1}$. $k^{\text{uni}}$ is the unimolecular rate constant given in s$^{-1}$. $T$ is
the temperature given in K.}
\label{tab:tab5}
\begin{tabular}{ccc}
\toprule
T & $\log k^{\text{bi}}$ & $\log k^{\text{uni}}$ \\
\midrule \midrule
\multicolumn{3}{c}{C$_2$}\\
\midrule
150&    -16.74& 6.92\\
120&    -17.58& 6.27\\
100&    -18.16& 5.87\\
90&     -18.46& 5.68\\
80&     -18.70& 5.57\\
70&     -18.98& 5.47\\
60&     -19.21& 5.47\\
\midrule 
\midrule
\multicolumn{3}{c}{C$_3$}\\
\midrule 
150&    -15.02& 7.84\\
120&    -15.69& 7.35\\
100&    -16.20& 7.00\\
90&     -16.43& 6.87\\
80&     -16.70& 6.72\\
70&     -16.92& 6.64\\
60&     -17.13& 6.62\\
\bottomrule 
\end{tabular}
}
\end{table}
\begin{table}
\parbox[t]{.45\linewidth}{
\centering
\caption{Instanton rate constants for the hydrogen abstraction from propargyl alcohol. $k^{\text{bi}}$
is the bimolecular
rate constant given in cm$^3$s$^{-1}$. $k^{\text{uni}}$ is the unimolecular rate constant given in s$^{-1}$. $T$ is
the temperature given in K.}
\label{tab:tab6}
\begin{tabular}{ccc}
\toprule
T & $\log k^{\text{bi}}$ & $\log k^{\text{uni}}$ \\
\midrule \midrule
\multicolumn{3}{c}{C$_1$}\\
\midrule 
300&    -13.49& 8.62\\
250&    -13.99& 8.23\\
200&    -14.67& 7.68\\
170&    -15.16& 7.29\\
150&    -15.51& 7.03\\
120&    -16.06& 6.62\\
100&    -16.43& 6.38\\
90&     -16.62& 6.26\\
80&     -16.84& 6.13\\
70&     -17.01& 6.07\\
60&     -17.25& 5.95\\
\midrule 
\midrule
\multicolumn{3}{c}{O}\\
\midrule 
350&    -17.79& 4.04\\
300&    -18.82& 3.10\\
250&    -20.11& 1.92\\
200&    -21.65& 0.51\\
170&    -22.68& -0.43\\
150&    -23.44& -1.11\\
120&    -24.77& -2.31\\
100&    -25.93& -3.37\\
90&     -26.65& -4.02\\
80&     -27.49& -4.79\\
70&     -28.35& -5.57\\
60&     -29.19& -6.31\\
\bottomrule 
\end{tabular}
}
\hfill
\parbox[t]{.45\linewidth}{
\centering
\caption{Instanton rate constants for the hydrogen addition to propenal. $k^{\text{bi}}$
is the bimolecular
rate constant given in cm$^3$s$^{-1}$. $k^{\text{uni}}$ is the unimolecular rate constant given in s$^{-1}$. $T$ is
the temperature given in K.}
\label{tab:tab7}
\begin{tabular}{ccc}
\toprule
T & $\log k^{\text{bi}}$ & $\log k^{\text{uni}}$ \\
\midrule \midrule
\multicolumn{3}{c}{C$_1$}\\
\midrule
200&    -17.42& 5.90\\
150&    -18.69& 4.87\\
120&    -19.48& 4.28\\
100&    -19.99& 3.96\\
90&     -20.25& 3.82\\
80&     -20.50& 3.72\\
70&     -20.71& 3.70\\
60&     -20.99& 3.67\\
\midrule\midrule 
\multicolumn{3}{c}{C$_2$}\\
\midrule
150&    -16.46& 6.72\\
120&    -17.22& 6.13\\
100&    -17.77& 5.73\\
90&     -18.00& 5.60\\
80&     -18.29& 5.42\\
70&     -18.50& 5.35\\
60&     -18.85& 5.19\\
\midrule 
\midrule
\multicolumn{3}{c}{C$_3$}\\
\midrule 
110&    -14.34& 8.97\\
100&    -14.58& 8.88\\
90&     -14.85& 8.75\\
80&     -15.13& 8.61\\
70&     -15.38& 8.51\\
60&     -15.66& 8.42\\
\midrule 
\midrule
\multicolumn{3}{c}{O}\\
\midrule 
290&    -16.20& 6.65\\
250&    -16.74& 6.21\\
200&    -17.57& 5.56\\
150&    -18.48& 4.88\\
120&    -19.06& 4.53\\
100&    -19.45& 4.35\\
90&     -19.69& 4.24\\
80&     -19.81& 4.28\\
70&     -19.96& 4.33\\
60&     -20.21& 4.35\\
\bottomrule 
\end{tabular}
}
\end{table}
\begin{table}
\parbox[t]{.45\linewidth}{
\centering
\caption{Instanton rate constants for the hydrogen abstraction from propenal. $k^{\text{bi}}$
is the bimolecular
rate constant given in cm$^3$s$^{-1}$. $k^{\text{uni}}$ is the unimolecular rate constant given in s$^{-1}$. $T$ is
the temperature given in K.}
\label{tab:tab8}
\begin{tabular}{ccc}
T & $\log k^{\text{bi}}$ & $\log k^{\text{uni}}$ \\
\midrule \midrule
\multicolumn{3}{c}{C$_1$}\\
\midrule
300&    -13.30& 9.14\\
250&    -13.81& 8.74\\
200&    -14.51& 8.19\\
170&    -15.03& 7.78\\
150&    -15.38& 7.51\\
120&    -15.97& 7.08\\
100&    -16.39& 6.79\\
90&     -16.63& 6.64\\
80&     -16.82& 6.54\\
70&     -17.68& 5.81\\
60&     -17.35& 6.28\\
\bottomrule
\end{tabular}
\newline
\newline
\centering
\caption{Instanton rate constants for the hydrogen addition to allyl alcohol. $k^{\text{bi}}$
is the bimolecular
rate constant given in cm$^3$s$^{-1}$. $k^{\text{uni}}$ is the unimolecular rate constant given in s$^{-1}$. $T$ is
the temperature given in K.}
\label{tab:tab9}
\begin{tabular}{ccc}
\toprule
T & $\log k^{\text{bi}}$ & $\log k^{\text{uni}}$ \\
\midrule \midrule
\multicolumn{3}{c}{C$_2$}\\
\midrule
150&    -18.34& 6.84\\
120&    -19.70& 6.31\\
100&    -20.93& 5.80\\
90&     -21.56& 5.63\\
80&     -22.29& 5.47\\
70&     -23.15& 5.34\\
60&     -24.25& 5.22\\
\midrule 
\midrule
\multicolumn{3}{c}{C$_3$}\\
\midrule 
120&    -14.64& 8.67\\
100&    -15.04& 8.47\\
90&     -15.26& 8.36\\
80&     -15.52& 8.25\\
70&     -15.76& 8.18\\
60&     -16.03& 8.16\\
\bottomrule 
\end{tabular}
}
\hfill
\parbox[t]{.45\linewidth}{
\centering
\caption{Instanton rate constants for the hydrogen abstraction from allyl alcohol. $k^{\text{bi}}$
is the bimolecular
rate constant given in cm$^3$s$^{-1}$. $k^{\text{uni}}$ is the unimolecular rate constant given in s$^{-1}$. $T$ is
the temperature given in K.}
\label{tab:tab10}
\begin{tabular}{ccc}
\toprule
T & $\log k^{\text{bi}}$ & $\log k^{\text{uni}}$ \\
\midrule \midrule
\multicolumn{3}{c}{C$_1$}\\
\midrule 
300&    -14.48& 7.72\\
250&    -15.15& 7.17\\
200&    -16.01& 6.45\\
170&    -16.65& 5.91\\
150&    -17.08& 5.56\\
120&    -17.79& 5.00\\
100&    -18.22& 4.70\\
90&     -18.43& 4.57\\
80&     -18.88& 4.21\\
70&     -19.14& 4.06\\
60&     -19.28& 4.06\\
\midrule 
\midrule
\multicolumn{3}{c}{O}\\
\midrule 
350&    -16.98& 5.35\\
300&    -17.79& 4.63\\
250&    -18.72& 3.81\\
200&    -19.70& 2.96\\
170&    -20.32& 2.45\\
150&    -20.74& 2.10\\
120&    -20.42& 2.55\\
100&    -20.48& 2.60\\
90&     -20.98& 2.17\\
80&     -21.09& 2.14\\
70&     -21.07& 2.27\\
60&     -21.24& 2.22\\
\bottomrule 
\end{tabular}
}
\end{table}

\begin{table}
\parbox[t]{.45\linewidth}{
\centering
\caption{Instanton rate constants for the hydrogen addition to propanal. $k^{\text{bi}}$
is the bimolecular
rate constant given in cm$^3$s$^{-1}$. $k^{\text{uni}}$ is the unimolecular rate constant given in s$^{-1}$. $T$ is
the temperature given in K.}
\label{tab:tab11}
\begin{tabular}{ccc}
\toprule
T & $\log k^{\text{bi}}$ & $\log k^{\text{uni}}$ \\
\midrule \midrule
\multicolumn{3}{c}{C$_1$}\\
\midrule
190&    -17.08& 6.15\\
170&    -17.52& 5.79\\
150&    -18.04& 5.37\\
120&    -18.91& 4.68\\
100&    -19.56& 4.18\\
90&     -19.90& 3.95\\
80&     -20.17& 3.80\\
70&     -20.53& 3.60\\
60&     -20.87& 3.47\\
\midrule 
\midrule
\multicolumn{3}{c}{O}\\
\midrule 
310&    -16.71& 5.61\\
270&    -17.34& 5.07\\
230&    -18.13& 4.38\\
190&    -18.97& 3.67\\
170&    -19.39& 3.32\\
150&    -19.77& 3.03\\
120&    -20.40& 2.57\\
100&    -20.76& 2.35\\
90&     -20.95& 2.25\\
80&     -21.09& 2.22\\
70&     -21.29& 2.15\\
60&     -21.61& 2.01\\
\bottomrule 
\end{tabular}
}
\hfill
\parbox[t]{.45\linewidth}{
\centering
\caption{Instanton rate constants for the hydrogen abstraction from propanal. $k^{\text{bi}}$
is the bimolecular
rate constant given in cm$^3$s$^{-1}$. $k^{\text{uni}}$ is the unimolecular rate constant given in s$^{-1}$. $T$ is
the temperature given in K.}
\label{tab:tab12}
\begin{tabular}{ccc}
\toprule
T & $\log k^{\text{bi}}$ & $\log k^{\text{uni}}$ \\
\midrule \midrule
\multicolumn{3}{c}{C$_1$}\\
\midrule
290&    -12.93& 9.53\\
250&    -13.25& 9.30\\
200&    -14.04& 8.82\\
170&    -14.34& 8.47\\
150&    -14.69& 8.20\\
120&    -15.29& 7.77\\
100&    -15.74& 7.46\\
90&     -15.95& 7.34\\
80&     -16.16& 7.23\\
70&     -16.39& 7.13\\
60&     -16.58& 7.10\\
\midrule 
\midrule
\multicolumn{3}{c}{C$_2$}\\
\midrule 
330&    -14.44& 7.64\\
290&    -14.87& 7.29\\
250&    -15.44& 6.82\\
200&    -16.30& 6.09\\
170&    -16.88& 5.61\\
150&    -17.29& 5.28\\
120&    -17.92& 4.80\\
100&    -18.39& 4.46\\
90&     -18.59& 4.33\\
80&     -19.44& 3.58\\
70&     -19.26& 3.86\\
60&     -19.57& 3.69\\
\midrule 
\midrule
\multicolumn{3}{c}{C$_3$}\\
\midrule 
290&    -18.13& 3.89\\
250&    -19.10& 3.01\\
200&    -20.49& 1.76\\
170&    -21.39& 0.96\\
150&    -22.01& 0.41\\
120&    -22.94& -0.37\\
100&    -23.59& -0.89\\
90&     -23.89& -1.13\\
80&     -24.18& -1.33\\
70&     -24.42& -1.48\\
60&     -24.44& -1.37\\
\bottomrule 
\end{tabular}
}
\end{table}

\end{appendix}
\end{document}